\documentclass[11pt,letterpaper]{article}

\RequirePackage[letterpaper]{geometry}
\RequirePackage{color}
\usepackage{graphicx}
\usepackage{authblk}
\usepackage[font=small,labelfont=rm]{caption}

\def\opex{ Opt.\ Express }
\def\apl{ Appl.\ Phys.\ Lett.\ }

\def\josab{ J.\ Opt.\ Soc.\ Am.\ B }
\def\nat{ Nature (London) }

\def\prb{ Phys.\ Rev.\ B }
\def\prl{ Phys.\ Rev.\ Lett.\ }

\title{Electromechanical tuning of vertically-coupled photonic crystal nanobeams}

\author{L. Midolo$^{1,3,*}$, S. N. Yoon$^{1,3}$, F. Pagliano$^1$, T. Xia$^1$, F. W. M. van Otten$^1$, M. Lermer$^2$, S. H\"{o}fling$^2$ and A. Fiore$^{1}$\\
\footnotesize 
\it $^1$COBRA Research Institute, Eindhoven University of Technology, P.O. Box 513, NL-5600MB Eindhoven, The Netherlands \\
\it $^2$Technische Physik and Wilhelm Conrad R\"{o}ntgen Research Center for Complex Material Systems, Universit\"{a}t W\"{u}rzburg, Am Hubland, D-97074 W\"{u}rzburg, Germany\\
\it $^3$These authors contributed equally to this work.\\
\textcolor{blue}{\underline{l.midolo@tue.nl}}
\normalsize \rm
}
\date{}

\begin{document}
\maketitle

\begin{abstract}
We present the design, the fabrication and the characterization of a tunable one-dimensional (1D) photonic crystal cavity (PCC) etched on two vertically-coupled GaAs nanobeams. A novel fabrication method which prevents their adhesion under capillary forces is introduced. We discuss a design to increase the flexibility of the structure and we demonstrate a large reversible and controllable electromechanical wavelength tuning ($> 15$ nm) of the cavity modes. 
\end{abstract}

\section{Introduction}
The use of Nano Opto Electro Mechanical Systems (NOEMS) for the real-time spectral reconfiguration of a photonic crystal cavity (PCC) has recently drawn a lot of attention because of the large achievable tuning rates with small optical losses [1--3] and of the possibility of application in tunable filters \cite{Chew10} and cavity optomechanics \cite{Winger11}. Moreover, a tunable PCC coupled to quantum dots (QDs) facilitates the spectral alignment of the cavity to the emitter, opening up the opportunity to realize efficient, on-chip and scalable single photon sources \cite{Balet07} and to study cavity quantum electrodynamics phenomena \cite{Khitrova06}.
Most of these systems are usually based on the electromechanical control over the evanescent coupling of two almost identical semiconductor cavities. 
In a previous work \cite{Midolo11} we demonstrated a tunable InGaAsP double-membrane PCC whose resonant wavelengths could be electrostatically controlled over a 10 nm range. 
The use of vertically coupled cavities allows the simultaneous tuning of the cavity and the additional electrical control over active layers, located in one of the two membranes, e. g. by Stark tuning.
This article describes the design, the fabrication and the characterization of a tunable 1D nanobeam cavity on GaAs with embedded InAs QDs. 
1D cavities have various advantages compared to 2D cavities such as the smaller mode volumes, higher quality factors Q, higher compactness and ease of design \cite{Deotare09,Zain08}. 
Electrostatically tunable 1D PCCs have been already demonstrated using in-plane actuation of laterally-coupled nanobeams \cite{Frank10,Perahia10} while the vertically-coupled configuration and the electrostatic tuning of nanobeam cavities on GaAs has not been reported yet. We achieve electromechanical tuning over 15 nm with 15 V applied voltage.

\section{Design and theory of 1D PC nanobeams}
\label{sec:design}
\begin{figure}[htbp]
\centering\includegraphics[width=12cm]{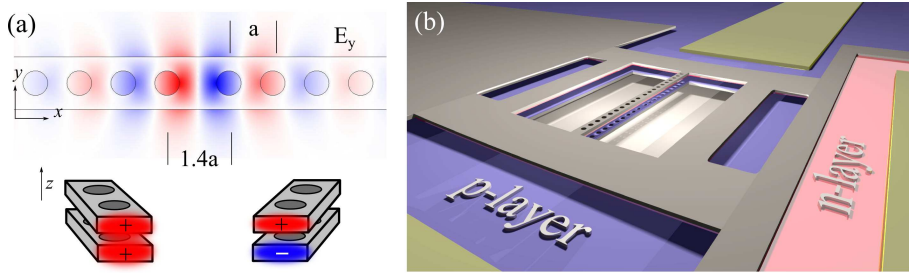}
\caption{\label{fig:theory}(a) Geometry of the 1D PCC on nanobeams and simulated in-plane mode profile of the $E_y$ component. The symmetric (left) and anti-symmetric (right) vertical profiles of the coupled system is also shown. (b) Sketch of the proposed diode structure to realize short and tunable nanobeams: a $12 \times 12 \mu\textrm{m}^2$ doubly-clamped bridge with a $8 \mu\textrm{m}$ long nanobeam in the center. Only the holes and the side trenches are etched through both membranes. By operating the junction under reverse bias, the electrostatic force bends the upper slab and brings the nanobeams at a closer distance.}
\end{figure}

The nanobeam PCC discussed in this work consists of a row of evenly spaced holes (lattice constant $a$) etched in a semiconductor beam of width $w$ and rectangular cross-section. By increasing one of the hole-to-hole spacing to $1.4a$ a cavity is formed. Such a simple design provides a single-mode cavity \cite{Joannopoulos} and it has been chosen for its simplicity. Fig. \ref{fig:theory}(a) shows the in-plane TE-like mode profile ($E_y$ component) obtained by solving Maxwell equations in two dimensions numerically.
In the configuration studied here, two nanobeams, one on top of the other, are brought at a very close distance ($<200$ nm) to obtain optical coupling. This results in an energy splitting of the y-polarized mode into two modes having an anti-symmetric (at higher energy) and a symmetric (at lower energy) vertical profile (Fig. \ref{fig:theory}(a)). By controlling the distance between the beams, the coupling strength can be modulated, resulting in a wavelength tuning of the cavity. 
To estimate the amount of coupling as a function of the geometrical parameters, a three-dimensional (3D) finite element method (FEM) is used \cite{Comsol} with a geometry adapted from \cite{Romer07}. 
By taking advantage of the symmetries in the geometry, only one eighth of the double nanobeam structure is simulated applying adequate boundary conditions on the symmetry planes. By enforcing a symmetric electric field (perfect magnetic conductor) or a symmetric magnetic field (perfect electric conductor) on the $z$-direction, the symmetric or the anti-symmetric modes can be calculated. A perfectly matched layer is used to simulate open boundaries.
3D simulations can be directly compared to photoluminescence (PL) experiments by integrating the radiated power (Poynting vector) from a dipole source \cite{Xu98} in the cavity to the surrounding air domain, assuming an ideal objective, capable of collecting light emitted in all directions.
\begin{figure}[htbp]
\centering\includegraphics[width=13cm]{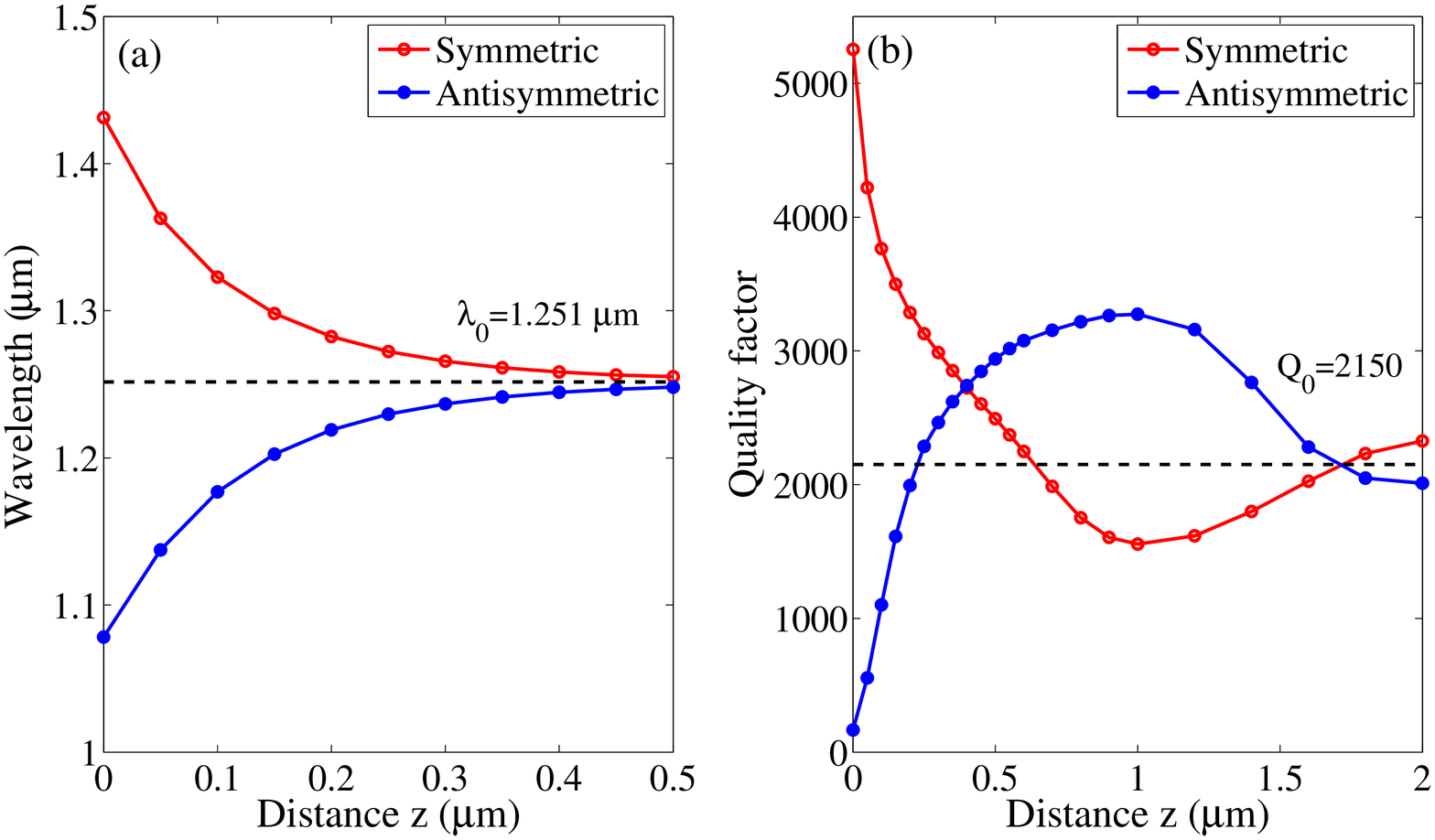}
\caption{\label{fig:sim} (a) Results of a 3D FEM simulation of double nanobeams. The peak wavelength of the symmetric and the anti-symmetric modes are plotted against the intermembrane distance. The parameters are: thickness $t=160$ nm, width $w=420$ nm, lattice spacing $a=370$ nm and hole radius $r=96$ nm. The refractive index dispersion in GaAs is taken into account. $\lambda_0$ is the wavelength in the uncoupled case (b) Calculated Q factor, obtained from the solution of the lossy eigenvalue problem. Q$_0$ is the Q factor in the uncoupled case.}
\end{figure}  
Fig. \ref{fig:sim}(a) shows the calculated cavity resonant wavelengths as a function of the nanobeam distance. A large shift ($\Delta\lambda=180$ nm) from the uncoupled nanobeam case is achievable when the distance goes to zero. At the nominal distance chosen for the experimental realization of the device ($z_0=200$ nm), a maximum tuning before pull-in (i.e. at $|z-z_0|/z_0 = 1/3)$ of $\Delta\lambda=23$ nm and a tuning rate $d\lambda/dz = 0.2$ nm/nm are predicted. The Q factor plotted in Fig. \ref{fig:sim}(b) is obtained solving an eigenvalue problem linearized around the mode wavelength. In the coupled region ($z<500$ nm) the symmetric mode shifts to lower frequencies, away from the light cone, and consequently its Q increases. Conversely, the Q of the anti-symmetric mode drops almost to zero. The Q behavior, apparently in contrast with the double slab case \cite{Notomi06}, is due to the Bragg mirror design (without tapers) which provides an abrupt termination of the cavity field and therefore a large amount of leaky components and radiation losses \cite{Deotare09,Akahane03}. As the nanobeams are moved further apart, a modulation of the Q is observed even without mode coupling. This can be explained by interference effects which cause an amplification or a cancellation of the leaky modes \cite{Thomas11}.
 
To complete the NOEMS model, an electro-mechanical analysis of the device is performed. An important mechanical design parameter is the overall stiffness of the NOEMS represented by the Hooke's spring constant per unit area $k$ (assuming a lumped electrostatic model made of metallic capacitor plates connected to springs). Taking the center of a doubly-clamped beam as the single degree of freedom of our model, the spring constant is given by $k=32Et^3/L^4$ where $E$ is the Young modulus ($E = 85.9$ GPa for GaAs), $t$ and $L$ are the nanobeam thickness ($t=160$ nm) and length, respectively \cite{Roark}. Assuming an uniform electric field, it is possible to derive the electrostatic force per unit area acting on each beam: $P_{electrostatic} = \epsilon_0 \frac{U^2}{2z^2}$ where $z$ is the distance between them and $U$ is the total applied bias. 
Applying the equilibrium condition to the symmetric system where two identical nanobeams are moving provides the equation $2P_{electrostatic} = k(z_0-z)$, $z_0$ being the distance at rest. The displacement curve and pull-in voltages can be calculated solving: 
\begin{equation}
z^3 -z_0z^2 + \frac{\epsilon_0 U^2}{k}=0
\end{equation}
A reasonable choice for $k$ is such that pull-in occurs at $U < 10$ V, which is also a typical breakdown value for our p-i-n diodes. This restricts the stiffness of the nanobeams to $k < 800$ Pa/nm and consequently sets a lower bound to their length ($L > 11 \mu\textrm{m}$).
As it will be discussed in section 3, it is possible to fabricate even longer structures. However a very large bending after undercut is observed (Fig. \ref{fig:fab}(c)) leading to unpredictable coupling configurations and altering the mechanical properties of the nanostructure. 
To keep the nanobeams shorter and, at the same time, to lower the demand of actuation voltage in the devices, the nanobeams can be mounted on a larger and more flexible frame structure which allows to rigidly translate the upper beam to the fixed bottom one. The proposed device geometry is shown in Fig. \ref{fig:theory}(b). Such a structure is not only more flexible but it also guarantees a more uniform application of the electric field than the simple nanobeam geometry. Due to the complexity of the shape, an effective stiffness $k_{eff}$, which can be calculated via FEM simulations or extracted from the measurements, is introduced. Since only the upper slab moves under the electrostatic pressure, the mechanical displacement corresponds to the distance between the nanobeams ($z$) and it is described by Eq. 1 replacing the stiffness by $k=2k_{eff}$. To satisfy the conditions discussed above, an ideal value of $k_{eff}$ should be lower than 400 Pa/nm. From FEM simulations, the structure stiffness ranges from 0.1 to 1 kPa/nm for the experimentally realized devices, depending on the actual geometry and dimensions.    

\section {Sample fabrication}
The sample is grown by molecular beam epitaxy on an undoped (100) GaAs substrate. A thick ($1 \mu\textrm{m}$) Al$_{0.7}$Ga$_{0.3}$As sacrificial layer is initially deposited to separate the double-membrane structure from the substrate. Then, two GaAs membrane layers having the same thickness (160 nm) and an Al$_{0.7}$Ga$_{0.3}$As inter-membrane layer (200 nm), as verified by scanning electron microscope (SEM) analysis, are grown. The upper membrane contains a layer of low density self-assembled InAs quantum dots, emitting around $1.3 \mu\textrm{m}$ at room temperature \cite{Alloing05}. To realize the electrostatic actuator, part of the membranes are doped to form a p-i-n junction. The top 50 nm of the lower membrane are p-doped and the bottom 50 nm of the upper membrane are n-doped ($n=p=3\cdot10^{18} \textrm{cm}^{-3}$). Since the QDs are situated above the junction, they are not affected by the electrostatic field.

\begin{figure}[htbp]
\centering\includegraphics[width=12cm]{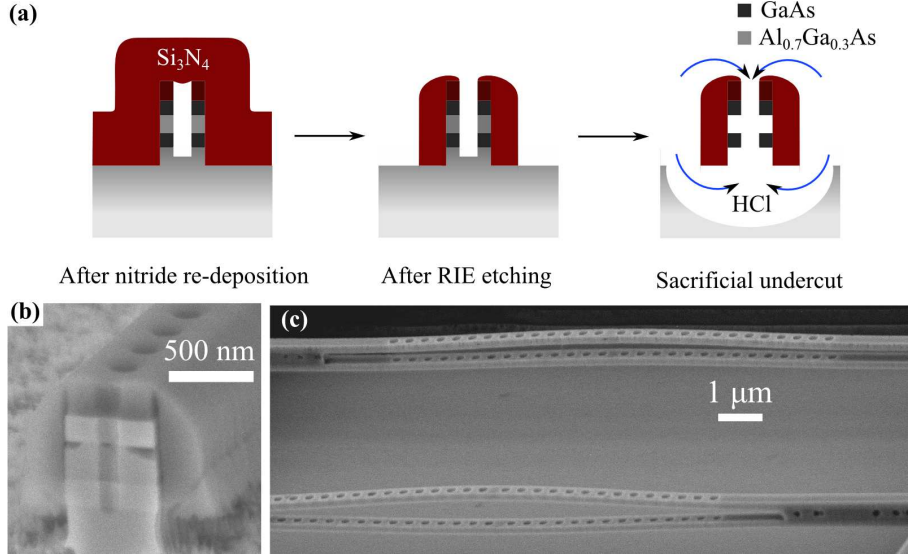}
\caption{\label{fig:fab} (a) The fabrication process used to realize freestanding nanobeams (seen in cross-section). (b) SEM picture of the cleaved hole cross-section after the sidewall fabrication but before wet undercut. The sidewalls cover the nanobeam but the holes are opened again. No Si$_3$N$_4$ is visible inside the hole. (c) 15 $\mu$m long free-standing nanobeams. Stress relaxation induces a large bending in different directions which may also cause the nanobeams to touch and adhere.}
\end{figure}

To fabricate the double nanobeams, a 400-nm-thick Si$_3$N$_4$ hard mask is first deposited by plasma enhanced chemical vapor deposition (PECVD). The proximity-corrected nanobeam design, consisting of holes and side trenches, is patterned by a 30 keV electron beam lithography (RAITH-TWO 150) on a 360 nm thick electron beam resist (ZEP 520A), aligned to the [011] or [0-11] directions. After development, the nanobeams are first transferred to the underlying hard mask using pure CHF$_3$ reactive ion etching (RIE) and then deeply etched ($\approx 800$ nm) through both GaAs membranes by inductively coupled plasma (ICP) (Cl$_2$/N$_2$ chemistry at 200 $^\circ$C).
To remove the residual Si$_3$N$_4$ mask and the sacrificial layer, a selective wet etching step in hydrofluoric acid is usually done at this stage. By doing so, however, the nanobeams will pin together during the drying step because of the very strong capillary forces developed and the low elastocapillary number ($<1$) of the nanobeams in water \cite{Mastrangelo93}. 
In \cite{Midolo11} we reported a method to avoid stiction without resorting to supercritical drying. It relies on the use of the Si$_3$N$_4$ mask to stiffen the structure during the drying process. For nanobeams this method is not directly applicable because after ICP etching the mask supports only the top nanobeam leaving the other one free to collapse. Here we introduce a new technique which consists in the fabrication of nitride sidewalls around nanobeams before the undercut. The process steps are summarized in Fig. \ref{fig:fab}(a).
 
The sample is cleaned in oxygen plasma and dipped into diluted HF:H$_2$O (1:100) for 10 seconds to smoothen the bottom of the recesses without starting the AlGaAs undercut.
Then a second deposition of a 600-nm-thick Si$_3$N$_4$ conformal layer is performed. The nitride is not deposited inside the holes, due to their small size ($r<100$ nm).
Using CHF$_3$/O$_2$ RIE at high power, the nitride is etched with a strongly anisotropic profile. Since the etch rate is much higher in the vertical than in the horizontal direction, a 300-nm-thick Si$_3$N$_4$ supporting layer is left on the side of the nanobeams. By carefully optimizing the RIE times, the holes are opened again without damaging the GaAs. Fig. \ref{fig:fab}(b) shows a SEM picture of the device cross-section after this step. Once the sidewalls are fabricated and cleaned, no Si$_3$N$_4$ is observed inside the holes and the sample is ready for the undercut.

To remove the sacrificial layer selectively with respect to both GaAs and Si$_3$N$_4$, a cold ($1 ^\circ$C) HCl 36\% solution is used \cite{Sun09} followed by a fast (3 seconds) dip in HF 5\% to remove possible residues. The acid etches the AlGaAs through the top holes and the sides. Subsequently, the sample is rinsed in ultra-pure water and dried with nitrogen. The Si$_3$N$_4$ sidewalls hold the nanobeams laterally and thus prevent their adhesion during drying.
Finally, the nitride is removed isotropically in a low power CF$_4$ plasma to minimize surface damage and to release the structures.
The resulting free-standing nanobeams are shown in Fig. \ref{fig:fab}(c). 
On long ($>10 \mu\textrm{m}$) freestanding structures, the upper and the lower nanobeam relax and bend either upwards or downwards. This situation is not desirable, since it causes a non-reproducible coupling after fabrication and may also cause the nanobeams to stick together. For this reason shorter nanobeams are fabricated and placed on larger (hence more flexible) movable frame, as discussed above.

To realize tunable structures, the process described above is realized on a previously prepared p-i-n junction with metal contacts. The diode's fabrication consists in two lithographic steps followed by wet etching (in citric acid/peroxide and HF 1\%) to open vias to the p- and to the n-layer. Lift-off of Ti/Au 50/200 nm is used to realize the contact pads. As discussed above, the larger structure sketched in Fig. \ref{fig:theory}(b), which serves as a supporting frame for the nanobeams, is defined during the p-via etching. All the movable parts are opened with holes to facilitate the undercut. The final device is shown in Fig. \ref{fig:result}(a).

\section{Measurement and results}
\begin{figure}[htbp]
\centering
\includegraphics[width=10.3cm]{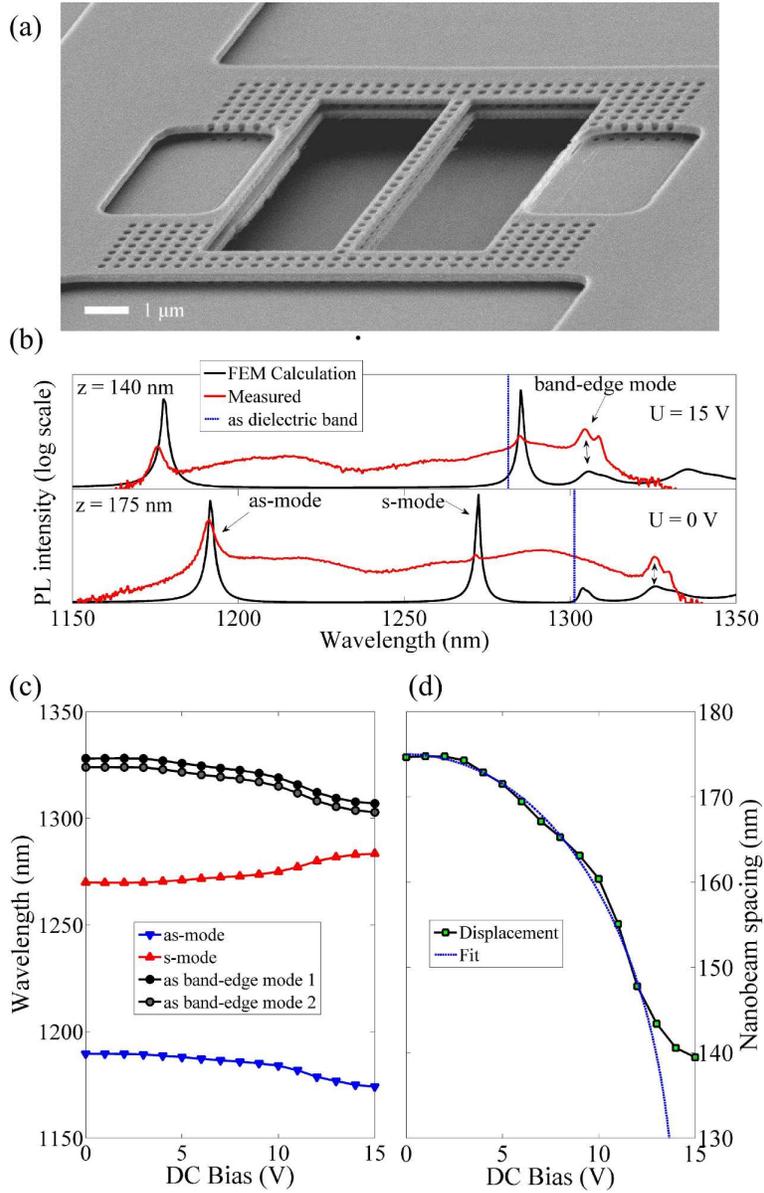}
\caption{\label{fig:result} (a) SEM image of the final device used for the tuning experiments. (b) PL spectra at 0 and 15 V DC bias (maximum tuning) compared to FEM simulations. (c) Tuning curve of the observed 1D PCC peaks, showing the as-mode, the s-mode and a double antisymmetric dielectric band-edge mode. (d) Calculated displacement of the upper membrane as a function of the voltage. The curve has been fit with a lumped electrostatic model to extract the equivalent device stiffness $k_{eff}$.}
\end{figure}
The device is tested with a micro-PL setup equipped with electrical probes. The QDs are pumped non-resonantly using a 785 nm diode laser from the top through a microscope objective (NA 0.4). The PL signal is collected from the objective, separated from the pump laser using a dichroic mirror, coupled into a fiber and analyzed with a spectrometer.  
The device is operated in reverse bias and for each voltage a spectrum is acquired.
The measurements reported here have been obtained with the device of Fig. \ref{fig:result}(a). 
The actual geometry of the device has been measured by SEM. The results are summarized in Table \ref{tab:meas}.
\begin{table}[htbp]
\centering
\caption{\label{tab:meas} PCC and nanobeam geometry as measured from SEM. When not explicitly specified, an uncertainty on the measurement of $\pm10$nm is expected.}
\begin{tabular}{ r || c | c }
	\hline
 	Nanobeam thickness & $t$ & 150 nm \\
 	Nanobeam width & $w$ & 420 nm \\
 	Nanobeam length & $L$ & (8 $\pm$ 0.2) $\mu$m \\
 	PCC lattice constant & $a$ & 370 nm \\
 	Hole radius & $r$ & 96 nm \\
 	\hline
\end{tabular}
\end{table}

Fig. \ref{fig:result}(b) shows the comparison between the PL and the FEM-simulated spectra using the parameters of Table \ref{tab:meas} and different air gaps. A good agreement is observed for an initial gap $z_0=175$ nm and for the maximum displacement of 35 nm ($z = 140$ nm). The calculated initial gap is lower than the nominal thickness of the inter-membrane sacrificial layer because of the relaxation of the structure after undercut. 
As expected, the single-mode cavity shows a double peak due to the coupling of the nanobeams. The symmetric (s-)mode (initially at $\lambda=1270$ nm) and the antisymmetric (as-)mode (at $\lambda = 1190$ nm) shift to longer and shorter wavelengths with increasing bias, respectively. The maximum shift is $\Delta\lambda_{as} = -15.4$ nm for the as-mode and $\Delta\lambda_{s} = +13.6$ nm for the s-mode. The tuning curve as a function of the applied reverse bias is plotted in Fig. \ref{fig:result}(c). The measured Q factor is $\textrm{Q}_s = 740 \pm 40$ for the symmetric mode and $\textrm{Q}_{as} = 450 \pm 20$ for the anti-symmetric mode. For the range of displacement considered here, the tuning of the Q, predicted by simulations, is not visible whereas the expected difference between the Q factors ($\textrm{Q}_s > \textrm{Q}_{as}$) is observed. The low Q compared to simulations is attributed to the poor selectivity of CF$_4$ plasma towards GaAs which damages the holes and reduces the thickness of the upper nanobeam. The latter is also causing the small difference of tuning rates and PL intensity between the s and the as mode. In the PL spectra, two quasi-degenerate band-edge modes (matching the simulated data) are also visible at $\lambda=1326$ nm at 0 V. 
These modes arise from slow-light PL enhancement \cite{Viasnoff05} at the dielectric band edge and they are de-localized over the nanobeam length. Due to the double layer structure, they also split into s and as profiles. A higher tuning range compared to the cavity ($\Delta\lambda_{BE} = -21.2$ nm) is obtained. This can be explained considering that the electromagnetic field of the dielectric BE has an in-plane distribution which is mostly located in the high-index area of the GaAs nanobeam, therefore modulation of the effective index in these areas is expected to affect these modes more than the cavity mode. This has been also verified by 3D band calculations of the double photonic crystal nanobeam and the resulting shift of the anti-symmetric dielectric band has been plotted in Fig. \ref{fig:result}(b) (dashed line).

From the value of the coupling it is possible to estimate via FEM simulations the distance between the nanobeams and to derive the displacement as a function of the applied bias (Fig. \ref{fig:result}(c) squares). The curve has been fitted using the lumped model of Eq. 1 (in the case of fixed bottom membrane) to estimate the effective stiffness $k_{eff}$ of the entire device. A value of $k_{eff} = (1.1 \pm 0.1)$ kPa/nm is obtained. This value is still higher than the design parameters discussed in Section \ref{sec:design} and more flexible designs are needed to achieve even higher tuning ranges. The displacement curve also shows a large deviation from theory when the bias voltage is $>11$ V. The saturation is due to the breakdown of the p-i-n junction, causing the currents in the intrinsic region to increase and limiting the voltage across the air gap.

\section{Conclusions}
A GaAs NOEMS device for the wavelength tuning of a 1D PCC on nanobeams has been demonstrated. By designing a flexible structure, a large reversible shift of the anti-symmetric mode ($\Delta\lambda_{as} = 15.4$ nm) has been observed with an applied bias of $U=15$ V. A good agreement with FEM simulations has also been obtained. Higher Q cavities may be obtained applying more sophisticated PCC designs and improving the fabrication process. The 1D nature of nanobeams opens up several opportunities for the design of on-chip tunable filters. It also enables the realization of a mixed in-plane and out-of-plane actuator, combining up to four nanobeams, to increase the tuning even further. Moreover, the possibility to combine a tuning control with an active region, makes such a device highly attractive for the realization of tunable lasers, single photon sources and quantum photonic integrated circuits.
\section*{Acknowledgements}
We acknowledge useful discussions with H. P. M. M. Ambrosius, T. B. Hoang, R. W. van der Heijden and S. Keyvaninia. This research is supported by the Dutch Technology Foundation STW, applied science division of NWO, the Technology Program of the Ministry of Economic Affairs under project No. 10380, the FOM project No. 09PR2675, the State of Bavaria and NanoNextNL, a micro and nanotechnology program of the Dutch ministry of economic affairs, agriculture and innovation (EL\&I) and 130 partners.


\end{document}